# A Clinical Approach to Training Effective Data Scientists

Kit T Rodolfa
University of Chicago
Chicago, IL
krodolfa@uchicago.edu

Adolfo De Unanue
University of Chicago
Chicago, IL
adolfo@uchicago.edu

Matt Gee
University of Chicago
Chicago, IL
mattgee@uchicago.edu

Rayid Ghani
University of Chicago
Chicago, IL
rayid@uchicago.edu

## Abstract

Like medicine, psychology, or education, data science is fundamentally an applied discipline, with most students who receive advanced degrees in the field going on to work on practical problems. Unlike these disciplines, however, data science education remains heavily focused on theory and methods, and practical coursework typically revolves around cleaned or simplified data sets that have little analog in professional applications. We believe that the environment in which new data scientists are trained should more accurately reflect that in which they will eventually practice and propose here a data science master's degree program that takes inspiration from the residency model used in medicine. Students in the suggested program would spend three years working on a practical problem with an industry, government, or nonprofit partner, supplemented with coursework in data science methods and theory. We also discuss how this program can also be implemented in shorter formats to augment existing professional masters programs in different disciplines. This approach to learning by doing is designed to fill gaps in our current approach to data science education and ensure that students develop the skills they need to practice data science in a professional context and under the many constraints imposed by that context.

## Introduction

A century ago, whether a patient's doctor had received any practical experience during their medical schooling was a haphazard function of the whims of their instructors. According to Long [25]:

> Some professors required virtual indentured servitude, whereas the educational experiences offered by others were too brief to be of value. Indeed, young surgeons might never have performed an operation.

With the advent of the residency model, medical training not only became increasingly standardized, but also ensured young doctors gained critical practical experience working with patients and taking on escalating responsibility over time. Although residency has been established in medicine for so long no data seem to exist to speak to its impact, its adoption is generally credited with revolutionizing the training and competency of new doctors. Adoption of similar models for students in psychology and, more recently, education [1, 35] further illustrate the importance hands-on training in practical fields.

Just as the field of medicine reflects the practical applications of basic research in biology, physiology, and anatomy, the emerging field of data science is primarily concerned with the practical application of knowledge developed in the more theoretical fields of computer science, machine learning, and statistics. Unlike medicine, however, graduate education in data science remains heavily focused on these theoretical underpinnings.

For the purposes of this paper, we define data science as a broad set of computational and data-driven methodologies and techniques coming from computer science, statistics, math, social sciences including machine learning and artificial intelligence used to tackle real-world problems. Our audience here is students interested in learning how to solve problems using data science, and not develop new methods.

The extent and quality of students' work with practical, real-world data (if any) during their education is highly dependent on their particular degree program and often cited as an important gap in the current educational environment [3, 8, 20, 22]. While the cumulative exercises (or capstone projects) employed by some programs can provide students with a valuable introduction to the nuance and challenges of working with real data, some have voiced concerns that these are not consistently implemented [5, 14], and we contend that developing these practical skills and intuition requires sufficient time and experience that it should live at the core of data science degree programs.

As a result, many students entering industry or the public sector with a master's degree in data science may have strong foundations in understanding the technical details of a wide variety of methods and techniques, but little experience in choosing the right method for a given problem, applying it to messy, real-world data, solving a problem with real-world constraints, and explaining results of their work to decision makers and stakeholders. Deficits in communication abilities and relevant experience with practical problems were highlighted among the critical skills for new hires in an employer survey performed by an American Statistical Association working group on Master's education in statistics [2], while other researchers are advocating for the inclusion of more practical data science skills in the undergraduate and graduate curriculum [3, 4, 6, 11, 18, 19, 36].

As with other professional disciplines, these practical skills also need to be grounded in a strong foundation of professional ethics and an understanding of existing legal regulations around data collection and use. The history of medicine again provides some helpful context: in his seminal work on the development of medicine as a profession in American, Starr [37] described self-regulation and a code of ethics with an enshrined service orientation as defining characteristics of professional sovereignty. Regardless of whether they work in the public, private, or non-profit sector, data science practitioners will find themselves facing the legal and ethical implications of their work on a regular basis, whether in the form of requests from supervisors to focus on certain aspects of their data, the potential for disparate impact of their predictive models, or the uses and protection of individuals' sensitive data. The recent interest and active conversation around the nature of professional ethics for data scientists [13, 16, 29, 31] only serves to highlight how critical it is that students entering this field are well-equipped with a toolkit that helps them navigate these issues. Any program seeking to train data scientists would be remiss to treat professional ethics as anything less than a central component of their curriculum.

Recently, a number of innovative degree programs have begun to seek ways to incorporate deeper practical experience relevant to employers (including master's programs at Northwestern [26] and Boston University [28]) or broader cross-discipline curriculum (for instance, in a new data science institute and undergraduate program at UC San Diego [17]).

Here we extend on these ideas to propose a new model for graduate education in data science, drawing inspiration from the medical residency and centered around a prolonged applied project as the focal point of the degree. A degree program focused on gaining real-world experience will be better aligned with ensuring that students develop the skills they need to be successful data science practitioners as they begin their careers. While we believe coursework highlighting the theoretical basis of data science methods should remain an important part of a student's experience, 1) it should span several disciplines including computer science, statistics, and social sciences and 2) it should serve to supplement and enhance students' applied work rather than act as the primary element of the program.

## Domains of Data Science Competency

Recent work in medical [25] and psychology [32] education has begun to view professional education in these areas through the lens of core competencies, providing a valuable framework for designing degree and post-degree programs that best serve the needs of students. Table 1 presents the domains of data science competency we considered while developing our proposed program, with more detail on each domain below. While a given data scientist may focus their career on certain class of problems or methods, this list reflects a core set of foundational skills that cut across these areas of specialization.

Table 1: Core Domains of Data Science Competency

| **Domain** | **Key Skills** |
|---|---|
| Problem Definition | Defining problem scope, identifying goals, and choosing appropriate methods and tools |
| Data Preparation | Cleaning, integrating, structuring, and storing messy real-world data, dealing with missingness and imputation. |
| Modeling and Analysis | Foundational methods in statistics, machine learning, and social sciences |
| Evaluating Results | Choosing appropriate performance metrics, measuring generalization performance, model selection, experiment design |
| Professional Ethics | Ethical conduct in the use and presentation of data science methods, bias and fairness in modeling, privacy and protecting sensitive data |
| Communication | Building trust in data, understanding organizational needs, communicating about complex methods and results, visualization, persuading people to use what you've done |

**Problem Definition** The first competency domain is the ability to define and scope a problem effectively. In our experience, we have framed scoping as involving answers to four questions [10]: 1) Who are the stakeholders for the analysis at hand and what are their goals for the project? 2) How will the output of the analysis be used (does it support certain actions)? 3) What data (either internal or external) is available and/or needed for the project? And, 4) What tools and methods are best suited for approaching the problem and how will results of the analysis be validated?

**Data Preparation** Most real-world problems involve datasets that are far more messy than anything students will have encountered in traditional academic programs. Effective data scientists need to be able to understand the data at hand as well as its context and generative process. They should be experienced with ingesting, cleaning, integrating, and storage of raw data files from multiple sources, dealing with gaps and missingness in the data, and preparing the data for further analysis or modeling.

**Modeling and Analysis** Traditional degree programs are well-suited at building one aspect of this competency domain, which is developing the toolkit of modeling and analytical methods. We need to augment the theoretical understanding with some key skills:

First, they need the ability to apply these methods to practical problems. Professional data scientists are best equipped when they understand the applications of these methods, their theoretical underpinnings, and crucially their limitations. Successfully applying these methods also requires sufficient domain understanding, ability to communicate with experts in that area, and the experience to turn that into a set of features/predictors that need to be provided to the modeling methods.

Second, in addition to building a large number and variety of models, data scientists need the ability to select models that are likely to perform well[1] in the future. Since real-world data science problems can be approached with a variety of different methods, effective model selection is essential to good data science. The model selection process needs an understanding of not only what the models are but more critically, how they will be applied: what metrics are going to be used to evaluate them? When multiple approaches are available, how should these be compared relative to one another and the best option chosen for validation trials and eventual deployment?

**Evaluating Results** Successful data science projects must not only provide results that perform well on existing data, but which actually continue to perform as expected when put into practice. While we have techniques to do initial model evaluation using historical data, our true goal is to select a model that is effective in the future by answering some key questions. Does the model or analysis continue to work as expected based on its performance on the data used to develop it? If there are several candidates to put into use, how will their real world performance be compared for the purposes of choosing a final option to deploy. Validating a model or analysis in a real setting with truly novel data is a particularly important aspect of successful data science work, often combining machine learning methods with the design, execution, and analysis of a randomized experiment (or non-experimental pilot) as well as soliciting and integrating feedback from non-technical end users. Likewise, the responsibilities of a data scientist don't end when a model or analysis is deployed – they must also consider how a deployed system will be monitored for performance degradation and how outcomes of the system actually impact equity and fairness over time.

**Professional Ethics** In the context of data science, ethical conduct involves both technical and non-technical skills. Students of data science should receive training in how to approach problems through an ethical perspective and ensure results they present reflect a fair and well-supported reading of the data. They need to understand the ethics around the use of the data they have access to, and possible implications on people affected by the system they are building. Likewise, they should be prepared for difficult situations in which they encounter institutional or managerial pressure to present a less scrupulous read of the data. Training in understanding issues around transparency of the work being done and accountability of downstream impact is critical here. Effective data science professionals should also be well-versed in technical methods for measuring bias and fairness in their results (and underlying data), as well as considering the implications of trade-offs in how these terms are defined or measured. A working knowledge of policy and legal frameworks for protecting private or sensitive data as well as research involving human subjects is also critical to the effective and ethical practice of data science.

[1] Well = everything we care about: accurate, fair, interpretable, stable, etc.

**Communication** As with ethics, communicating about data science can involve technical and non-technical skills. Professional data scientists need to be able to understand needs of the organization in which they work and how those needs translate into data at hand (or that could be collected). Likewise, they need to be capable of building organizational trust in the use of analytics and data to drive decision making. A critical skill is the ability to effectively communicate about complex methods and results to non-technical audiences, often involving methods for data visualization and developing interpretable explanations from otherwise black-box models.

## Data Science Project Lifecycle

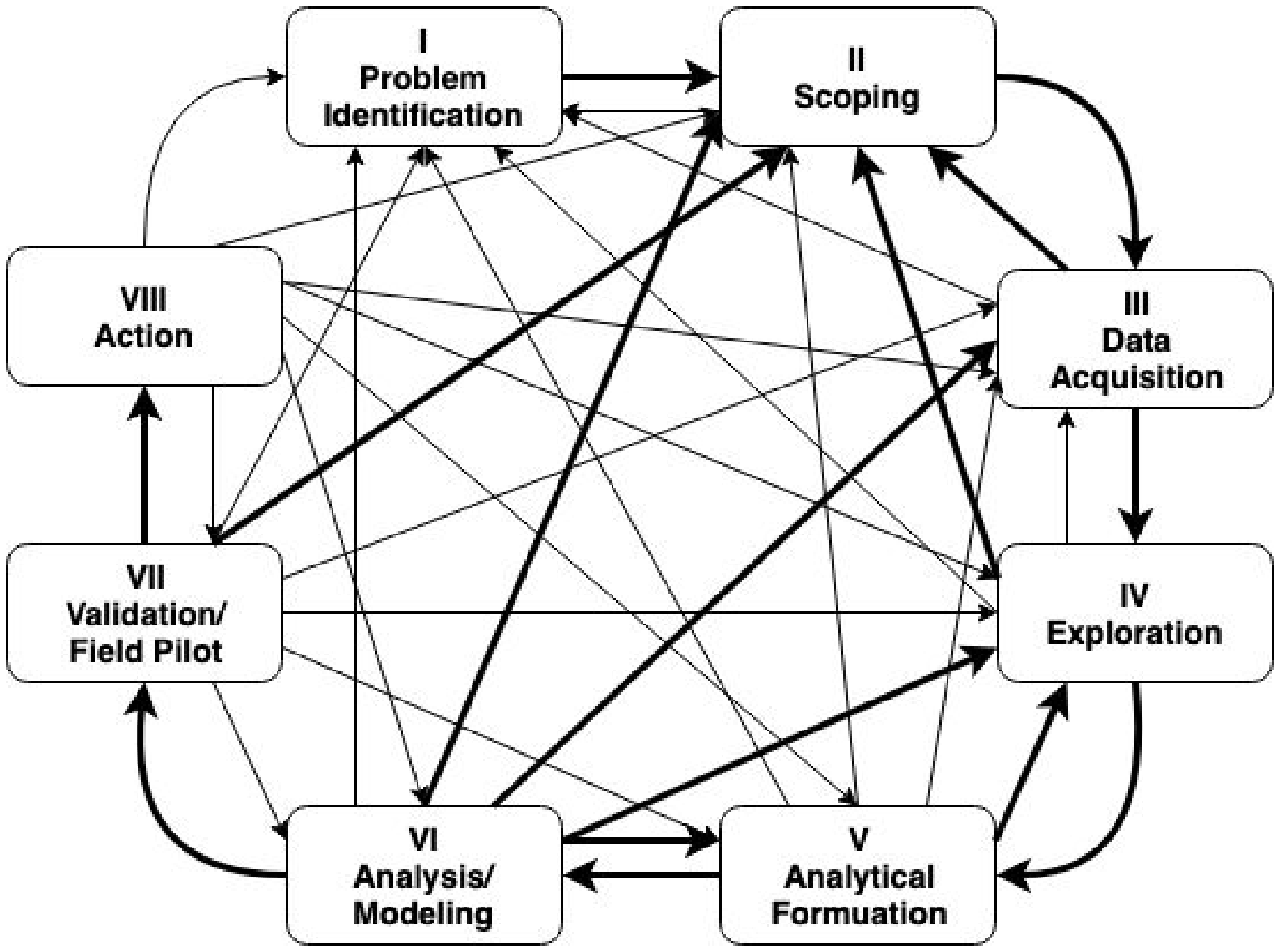

Figure 1: Lifecycle of a data science project. The phases of real projects are highly interconnected, with insights from later phases informing and revising earlier ones, resembling more of a “web” than a typical linear pipeline or cycle. Bold arrows show more frequent transitions typically repeated multiple times in the course of a project.

While the discussion above provides a structure for thinking about the competencies required for a data scientist to be effective, these domains are far from independent of one other and are used repeated over the course of a given project. Figure 1 shows the interconnectedness of the different phases of a typical data science project, while Figure 2 illustrates how skill domains are put into use across these phases.

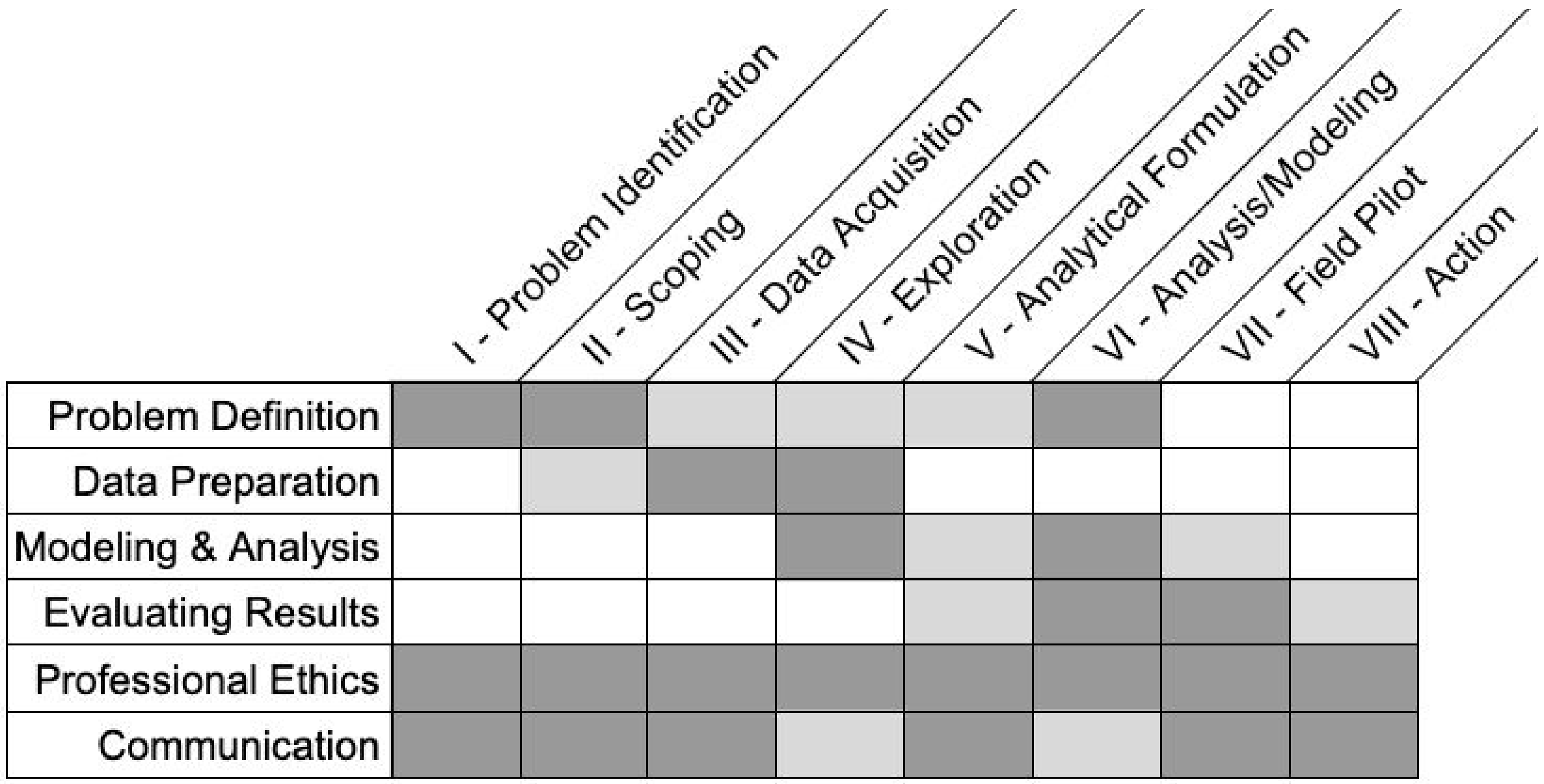

Figure 2: Competency domain use through the lifecycle of a data science project.

These phases are described in more detail below, illustrated through a real-world example of reducing lead poisoning in children:

**Phase I - Problem Identification** Even before the outset of a data scientist's work, stakeholders or policy makers may need to make a decision about pursuing a solution to a problem. This involves evaluating whether the problem is significant, whether it's feasible to solve the problem with data science, and whether there is (or will be) commitment internally to allocate resources to addressing the problem. Data Scientists have a critical role in this process, both providing a voice about what is technically feasible and why it may provide improvements over current practices, as well as an ethical duty to highlight the limitations and risks involved. Similarly, a solid understanding of laws and best practices around data privacy and sharing can be essential to helping decision makers understand how the data they have can be used and what other data they may or may not be able to collect for the project.

*Example: Public health officials identify high rates of lead poisoning in children in their jurisdiction, but current practice only remediates issues in homes after a child has tested positive for elevated blood lead levels. They would like to reduce lead poisoning in children by proactively identifying children who may be at risk before poisoning occurs.*

**Phase II - Scoping** As a project begins to get underway, competencies in communication and problem definition will be particularly important in scoping the actual work. The data scientist needs to be able to evaluate what questions can be answered with the available data as well as work closely with the stakeholders to understand their needs and how any models and analyses will actually be put into use [10]. Ethical concerns at this stage include considering how sensitive data will be handled and protected as well as establishing criteria by which analysis will be evaluated in ways that balance efficiency, effectiveness, and equity.

*Example: A scoping session is held including public health officials, clinicians, lead hazard inspections teams, and data scientists to understand the data available and how risk scores would be put into use. Because of the need to work with private health information and data pertaining to children, the decision is made to restrict all analytical work to the Department of Public Health's secure server environment. Primary intervention is identified as lead hazard inspections in homes with high risk of lead hazards and presence of a child under 12 months. The key goal identified in the scoping phase was to effectively reduce childhood lead poisoning in an equitable manner across underserved communities.*

**Phase III - Data Acquisition** Acquiring, storing, linking, understanding, and preparing data for analysis in a real-world project is often an involved and iterative process, requiring working closely with the owners of various data sources to ensure any transferred data is provided in a consistent and reliable format and necessary steps are taken to protect private or sensitive information. During this phase of work, the data scientist needs to apply skills working with and structuring raw data to get it into a storage format that is appropriate for linking it with other data sources. Each of those steps requires active communication with the project's stakeholders to understand the context in which the data was collected and structured, its idiosyncrasies, and ensure data definitions actually describe the events they are supposed to reflect.

*Example: The Department of Public Health provides a database and server for analysis in their environment with an extract of individual-level blood lead test results as well as inspection reports from lead hazard inspections. Data from additional sources are imported into the environment, including census data, childhood nutrition benefit program data (to identify potentially vulnerable children), and information about buildings from the county assessor website. Address normalization and geocoding allows data to be linked across these sources and data scientists work closely with the owner of each data source to ensure they understand the data structures and fields.*

**Phase IV - Exploration** This initial phase of analysis focuses on exploring the trends and relationships in the data through summary statistics, visualization, and preliminary modeling. Although many traditional data science programs will provide students with a basic toolkit of applicable statistical methods, the open-ended nature of working with messy real-world data can be daunting to students who have only worked with highly curated data sets in a guided setting. In most projects, this stage also requires a facility with handling missing data as well as identifying potential bias and disparities in labels and potential features.

*Example: The data scientists use a combination of descriptive statistics, bivariate correlations, spatial and temporal analysis to begin to understand the relationships in the data and its*

*limitations. Missing values in the childhood nutrition benefit dataset identify an error in the ETL process that is corrected with a new data extract, while a sharp decrease in the number of blood tests in data older than 17 years reflects a change in policy around testing that defines the limitation in historical training data.*

**Phase V - Analytical Formulation** This phase involves formulating our initial problem as a concrete analytical problem. In most cases, a greater understanding of the available data and its nuances will result in a greater understanding of the problem itself as well. During this phase, a data scientist will need to be able to effectively communicate preliminary results to stakeholders, including any limitations or shortcomings. At the end of this phase, the data scientists and stakeholders will have a set of design decisions to set up the technical framework for the project., From this more well-informed perspective, the project scope can be revisited and modified, which may in turn require more data collection, feature engineering, or exploratory analysis.

*Example: Drawing on what they learned in exploring the data, the data scientists work with the public health officials to formulate a classification problem at the address level using blood lead levels above a specific level as a training label. Monthly risk scores will be produced for every house with a child under the age of 12 months to correspond with the planning cycle of the department's housing inspection team and evaluated on the basis of precision (positive predictive value) among the top 250 highest-risk addresses, consistent with their monthly capacity for lead inspections, as well as the representativeness of underserved communities in the results.*

**Phase VI - Analysis/Modeling** Many projects will move through multiple rounds of exploration and refinement, iteratively approaching a final analytical phase as the problem definition and scope continue to evolve. While the competency domains essential to these later analytical phases are similar to those employed in the earlier exploratory work, the specific skills used here will tend to shift away from data description and more towards summarization, prediction, and/or extracting meaning. Generally, this phase involves generating a large number of models, analyses, or results followed by analysis to draw meaningful conclusions. In the case of predictive modeling, this might involve the process of model selection, balancing different performance and fairness metrics to arrive at a single model (or small menu of choices) to put into practice. For analytical projects, this phase may also involve telling a story from the available data, putting to use not only communication and data visualization skills but also ethical frameworks for how to summarize vast amounts of data in fair and meaningful ways.

*Example: The data scientists run a grid of thousands of model specifications, including several families of classifiers and hyperparameters. Based on its ability to balance precision in the top 250 and balance false omission rates across race and socioeconomic status, a random forest model was chosen to test in a field trial.*

**Phase VII - Field Validation/Pilot** The previous phase results in a final set of analysis results or models that are ready to be piloted or validated in a field trial. This phase of the project involves designing the trial to test the ongoing effectiveness and usability of the analysis. In some cases this may involve developing a randomized control trial to measure the causal impact of deploying a predictive model, while in others it may involve collecting feedback on how a report impacts decision making. In any case, this phase should focus on validating that the results of the analysis in fact continue to perform as anticipated when presented with truly novel data, including with respect to relevant fairness metrics. Likewise, gathering qualitative feedback from decision makers acting on the analysis is an important aspect of the field pilot.

*Example: A one-year field trial was developed, during which a random 50% of the 250 highest-risk addresses were inspected for the presence of lead each month, and remediated where hazards were found. The trial confirmed the performance of the model in identifying children at risk of poisoning because of the presence of lead in their homes as well as its representativeness across communities.*

**Phase VIII - Taking Action** Finally, for a data science project to successfully impact the decisions or actions of policy makers or stakeholders, results must be clearly and effectively communicated to these (generally non-technical) audiences along with recommended actions or a menu of choices. The ethical obligations of responsible data science practitioners at this stage reach far beyond avoiding the colloquial idea of "lying with statistics" to an awareness of the potential societal impact of their work. Any recommended course of action involves trade-offs (for instance, between optimizing for overall efficiency vs fairness across affected groups) and the data scientist performing these analyses may be the best-positioned individual to articulate the trade-offs associated with any potential action.

*Example: Although the number of households with lead issues remediated was too small to have a significant impact on the number of children diagnosed with lead poisoning during the trial period, calculations suggested that deploying the model could appreciably impact lead poisoning over the following decade. The Department of Public Health decided to move forward with putting it into practice, committing resources maintain and periodically refresh and re-evaluate the model.*

Taken alone, each phase of a project draws on a range of different competencies, highlighting the need for well-rounded skill development in data science education. Moreover, the heavily interconnected nature of the project phases illustrated in Figure 1 reflects the importance of agility in applying these skills throughout the entire project lifecycle. Effective data scientists need to be able to recognize when initial exploration or analyses might dictate a change in scope and responsible practitioners will surface new problems they identify in their work with the data as candidates for future work. With their heavy focus on teaching analytical methods or tools through classroom instruction in a linear, siloed manner, we feel that many current efforts to educate data scientists fall short of their mandate to produce well-rounded practitioners who are equipped to handle the nuance of problems they will encounter in their career.

# A Better Way: The Data Science Residency Master's Program

For many of the skill domains described above, classroom-based coursework is an inefficient and ineffective method of building student competency. Instead, we propose a three-year Data Science Residency Master's program structured around a core of applied project work with supported by lectures and workshops.

Ideally, projects should be sourced from a diverse set of industry, non-profit, and government partners and reflect a variety of problem types and methods in order to give students a range of options to explore and allow them to match with a project that best fits their interests. Projects should be structured to answer a practical question using the partner's actual data and involve the entire scope of project definition through analysis to field testing and deployment. We call them "partners" because they are collaborators in the entire process as opposed to just providing a problem or a data set.

Additionally, we believe that classroom instruction will be most effective if it is tightly coupled with problems students are currently facing in their projects, providing context, frameworks, and tools to help them approach issues that are already at the top of their mind. A number of programs in computer science [15, 33] and introductory courses in data science [34] have found advantageous effects to pairing instruction with real-world context through projects and we have likewise seen this sort of "priming" effect at play several times in working with students in the Data Science for Social Good summer program [9], informing our perspective on how learning and practice interact. Our proposed approach to instructional hours for Residency Master's therefore breaks from the typical concept of quarter- or semester-long topic-centric courses in favor of a concept of instruction that better reflects the skills students will put to use through the lifecycle of a data science project. This approach builds on the materials we have previously created for the Data Science for Social Good summer program [9, 21] as well as for the Coleridge Initiative classes in Applied Data Analytics [23].

Here, we envision shorter segments of instruction on each topic, closely tied to one another as well as to the project work that forms the core of the student's experience. Each segment could be taught by faculty who specialize in the topic at hand or faculty associated with the data science program itself, but in either case we believe it is critical to have both close coordination of classroom curriculum across these domains as well as direct ties between instruction and challenges students are likely to be facing in putting this work into practice.

Our proposed structure for such a program is shown in Figure 3. Early in the program, a heavier focus on classroom instruction will ensure all students have a strong foundation in tools and methods, while three "project sprints" will focus on different phases of the data science project lifecycle. In their second and third years, instruction will play a supportive role as students focus primarily on project work, with their responsibilities increasing with seniority, and culminating in an oral defense.

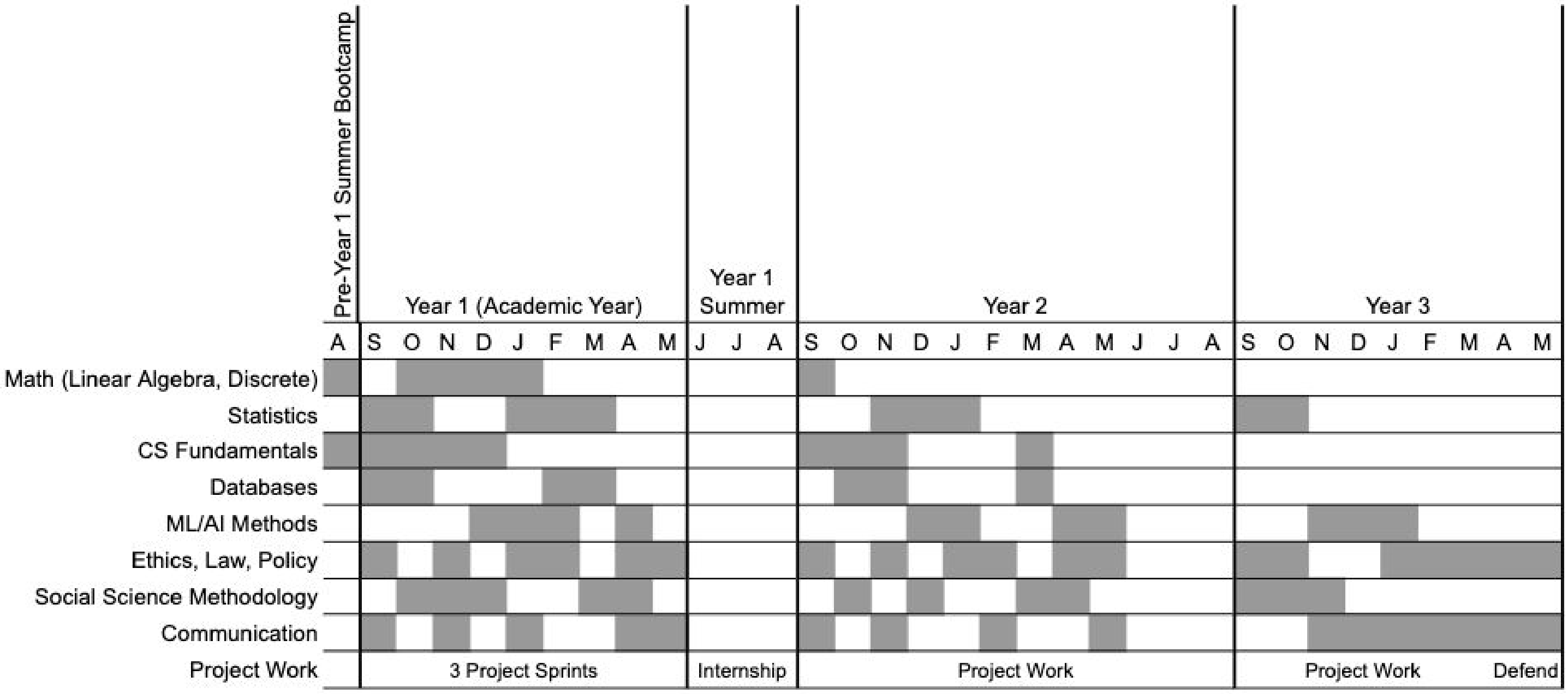

Figure 3: Proposed Distribution of Instructional Hours and Project Work for a Data Science Residency Master's Program

**First Year**

Prior to beginning the first year of their program, students would be expected to participate in (or, optionally, test out of) a bootcamp covering mathematics and programming fundamentals to ensure they begin the program with a common foundation. Following this summer introduction, students should be expected to have a basic understanding of linear algebra, probability, SQL, python, and programming workflows (editing & running code, bash, collaboration & version control, etc).

The program's first year is also heavier on coursework than the subsequent years in order to familiarize students with the full data science workflow and begin to develop their competency in each domain. Here, the first year is structured as a microcosm of the entire project lifecycle, supported by three "project sprints" developed from real-world problems and datasets (ideally drawn from the program's previous projects after it has been running for a few years):

- Sprint 1 focuses on the early phases of a project: evaluating need and convincing decision makers to support a project, problem definition and scoping, and data acquisition.

- Sprint 2 focuses on the more technical aspects of exploring the data and performing analyses as well as refining the project's goals and criteria as more knowledge is developed.

- Sprint 3 focuses on consolidating results, model selection, and turning results into action.

Each sprint would run for three months with tightly-coupled concurrent instruction. For instance, lectures in the domain of ethics, policy, and law during Sprint 1 might focus heavily on the existing legal and ethical practices for acquiring, using, and protecting sensitive data. During Sprint 2, lectures in this domain might focus on understanding and applying metrics to measure bias and fairness. And, in Sprint 3, they could focus on building frameworks for evaluating the potential societal implications of students' work and the inherent trade-offs involved. Likewise, instruction in statistics might evolve from focusing on data exploration early in the program to inference and model interpretability methods later.

During the summer following their first year, students would ideally participate in an internship at the partner organization for the project they will work on in their remaining two years. Doing so will allow them to form connections with the project partners as well as develop a working understanding of the organization's data and priorities.

**Second Year**

During the second year, the program's focus shifts to primarily emphasize project work. Classroom instruction should be limited to 3-6 hours per week, generally revisiting topics introduced in the first year with an increased level of depth and nuance. Just as instructional time in first year is coupled to the introductory project sprints, coursework in the second year should seek to provide the tools students will need in their ongoing projects just in time as they need them. Second year students' project work should begin with scoping and problem definition and shift towards a heavy focus on the technical aspects of understanding and analyzing the partner data after several months.

In our experience, students can also learn a great deal from each other in the course of their project work if given a forum in which to do so. Several types of interactions can be helpful here: First, frequent (e.g. weekly or bi-weekly) and highly structured status check-ins across all projects can provide context about how different organizations operate and move through phases of the work. Second, in-depth technical updates from each project on a less frequent basis (e.g., every other month) can provide exposure to methods and applications beyond a given student's project and give students the opportunity to provide feedback on each others' work. And, third, periodic focused skills sessions provide an opportunity to practice applying specific skills to other contexts, such as holding a scope-a-thon once a year to develop ideas for potential new projects based on their current work.

**Third Year**

The role and responsibilities of students should expand in their third year, including mentoring first and second year students as well as more direct interactions with project partners, both to communicate results and help develop the scope of future projects. Additionally, the regular cross-project check-ins and technical updates should continue to serve as a forum for exposing students to other problems, organizational contexts, and technical methods beyond their own projects. Formal coursework should again be fairly limited in the third year of the program, with an increased focus on practical curriculum around communicating about data science to non-technical audiences and understanding the role of data scientists in society.

Finally, we envision an oral defense as the culminating exercise of the student's program, allowing the student to demonstrate their mastery of the data science competencies and their practical application. In particular, we believe that it is critical for the defense committee to include representation from the project partner, giving them a voice in the student's final assessment (and that this expectation should be built into the partner agreement).

## Alternative Curriculum

While we believe the structure proposed above would best suit learning and retention by matching classroom instruction to the challenges students face at each phase of their project work, we also recognize the departure from traditional single-instructor, single-topic courses may be challenging to implement in some contexts. As such, Table 2 describes a structure that maintains the core elements of the Data Science Residency Master's but may fit more readily into existing settings.

Table 2: Alternative Structure for Data Science Residency Master's Program

| **Term** | **Coursework** | **Project Work** |
| --- | --- | --- |
| Pre-Year 1 Summer | Mathematics / Programming (Python and SQL) / Workflow Bootcamp | |
| Year 1 | Math (Linear Algebra and Discrete Math), Statistics, Computer Science Fundamentals (Data Structures and Algorithms), Databases, Machine Learning Methods, Social Science | 3 Project Sprints |
| Year 1 Summer | | Internship with Project Partner |
| Year 2 | Advanced Methods (ML, Causal Inference), Ethics/Fairness, 1 Elective | Project Work |
| Year 3 | Communication, Advanced Methods Elective (from ML, Stats, Operations Research, Social Sciences), 1 Elective | Project Work, Defense |

While this version of the program uses a more traditional structure for the coursework, we nevertheless believe it is important for this time in the classroom to be envisioned as primarily a means to support students' success in their practical work through real-world projects. As above, the first year pairs a heavier focus on classroom instruction with three project sprints introduce students to tools, methods, and theory they will need to apply to successfully, and responsibly, execute a data science project. Ideally, the sprints could be integrated with the assignments across these core courses. The second and third years allow students to focus more heavily on their project work with a considerably reduced course load, and the degree program culminates in an oral defense.

## Staffing and Structure

The real-world project work and external partnerships that are the central components of the Data Science Residency Master's have a number of implications for the structure and staffing the program will need in order to be successful.

First and foremost, the program will need to develop a pipeline of new projects through partnerships with government agencies, industry, and nonprofits, and will likely need staff who can cultivate and maintain these relationships. Maintaining a small team size for each project will be important to ensuring a well-rounded experience in which students participate in all aspects of the work (in practice, our experience with the Data Science for Social Good program might suggest that groups of more than 3 to 4 students can result in siloing). As such, the total number of projects will be an important determinant of the enrollment capacity of the program.

These external partnerships are critical to the success of the program overall. A good project partner is far more than a source of data for students to work with independently. Rather, they are an integral part of the project from end-to-end, committing resources and time to project scoping, extracting and explaining data, providing intermediate feedback and "gut checks", potentially executing on field trials, and participating in students' final evaluation. Moreover, most external partners will have several problems that can be developed into data science projects (and, often, many more to be discovered through the course of this work), so external partnerships should be envisioned as long-term relationships well worth investing programmatic resources in cultivating.

Second, the project work will need to be supported with human and technical infrastructure. This includes project managers who can help create and facilitate the ongoing relationships with the external partners through the course of the project as well as professionally managed servers and databases by a staff of devops engineers. These staff members can likely work across several projects & external partners, but the program will need to ensure both roles are staffed sufficiently to allow the project work to progress smoothly.

Third, a staff of full-time practicing data scientists could act as project leads and technical mentors, as well as providing a much-needed source of institutional knowledge about projects, tools, and datasets as students enter and leave the program. While institutional structure will dictate the specifics of the role (whether faculty, research faculty, senior fellow, or staff scientist), we envision it as a permanent position supported by hard money from the degree program and focused full-time on mentorship and supporting the applied data science work of the projects through practical research and tool development.

Structurally, the Data Science Residency Master's would ideally be envisioned as a collaboration that bridges several departments at an institution: computer science, statistics, public policy, government, medicine, public health, and potentially many others. While we recognize that creating a new cross-institution entity may pose a challenge for some institutions, there are several strategies and opportunities to foster broad participation and inclusion across existing departments and faculty [12, 24, 30, 36]. Existing data science programs might also consider adding the Residency Master's as an optional track or degree alongside their current curriculum. In addition to providing a smoother path to transition to a more clinical program over time, this could provide a good opportunity to study the outcomes of students who complete the two tracks to further refine the model and understand what additional skills they may need. Similarly, existing Masters in Public Policy, Public Administration, Social Work, and Public Health programs can add components of this model as a certificate or specialization to prepare professionals in those disciplines.

# Discussion

Current data science degree programs, with their heavy focus on coursework and theory, offer students little opportunity and training to undertake a significant project that reflects the messiness and constraints of working in a professional context. Viewed through the lens of competency domains that need to be mastered by effective practitioners, it becomes apparent that this current educational paradigm leaves significant gaps to be filled (one hopes) by their early years of experience in the workforce.

Here we have proposed a new approach to training data scientists, taking a cue from the concept of medical residency and putting the development of practical experience at the center of a student's education. This approach has been informed by our experience with designing and implementing different types of data science education programs including the Masters in Computational Analysis and Public Policy at the University of Chicago [27], the Data Science for Social Good Summer Fellowship [9], and the Applied Data Analytics for Governments program at the Coleridge Initiative [7] We believe this focus on learning by doing will develop more well-rounded data science graduates who are better equipped to begin their professional careers. At present, we are exploring opportunities to test this model, and encourage others to do the same, as well as experiment with variations on the program proposed here. As the discipline of data science continues to develop, much remains to be learned about best practices in training and education. We hope that the degree program model we have proposed will not only provide one step forward along that path of discovery but also spark other forms of innovation and experimentation aimed at training our students to have a positive practical impact through data science.